\documentclass[aps,pre,twocolumn,amsmath,amssymb,amsfonts]{revtex4}
\usepackage{epsfig}
\usepackage{graphicx}
\usepackage{dcolumn}
\usepackage{bm}

\begin{document}

\author{Alexander Jonathan Vidgop$^1$}
\author{Itzhak Fouxon$^{2}$}

\affiliation{$^1$ Am haZikaron Institute, Tel Aviv 64951, Israel}
\affiliation{$^2$ Raymond and Beverly Sackler School of Physics and Astronomy,
Tel-Aviv University, Tel-Aviv 69978, Israel}

\title{Web as a fundamental universal system}

\begin{abstract}

We study the possibility that further advancement in the understanding of the order of chaos may demand a certain reconsideration of the approach to the classical mechanics.
For this we suggest to consider the viewpoint that spatio-temporal relations between objects are emergent and that they are but a result of complex interactions of objects.
Such an approach is a natural continuation of the revision of the notions of space and time started by the general theory of relativity. It leads to
a possible extension of the structure of the classical mechanics. Namely, the result of the objects interactions can be wider than the spatio-temporal relations. In this case interactions form a generalized space ("Field-Space") wider than its spatio-temporal section (while the known interactions are embedded within that section). The study of such hypothesis demands constructing a theory, not relying on space and time as primitive notions. As primary elements of such a theory we consider objects and purely informational connections between them, not expressed in spatio-temporal terms. Following the logic of this theory, the world constitutes a complex information web. The Web is in the state of a constant flux (of a more general category than the one of the quantum fields) - an incessant change of connections, the complex order of which comprises the order of chaos. Reminding the space build-up from the Regge skeleton, the Web builds a unified Field-Space of pure information, manifested in space as energy distribution. We describe the emergence of spatio-temporal laws from the viewpoint of purely informational physics. The suggested construction of physics on the purely informational basis is a candidate for the theory of quantum gravity.

\end{abstract}

\maketitle

\section{Introduction: Space, Time and Chaos}

Physics developed from attempts to find order in the changes of spatial configurations of natural objects, such as celestial bodies. On the base of the classical mechanics, applicable in everyday situations, a picture emerged within which the relations between objects in space and time are a closed, self-explaining entity with future uniquely determined by the current spatial configuration of objects and the instantaneous rate of its change.

The subsequent discovery of chaos showed, however, that practically in every real situation the freedom of possible motions of the system, left after imposing the laws of classical mechanics, is extremely vast. The relations between objects in space and time that are realized in reality, remain largely inexplicable. In other words, in complex situations, the constraints constituted by the laws of mechanics are extremely loose and they bring little information, making us to search for additional connections. We are brought back to the original question - what is the order behind the changes of spatial configurations of objects around us? Can it be that the laws of mechanics are emergent from some more fundamental processes the understanding of which would shed new light on chaos? Is it possible that the spatio-temporal relations are but secondary and emerge on the basis of some deeper phenomena the essence of which has nothing to do with space and time?

Here it is important to notice the principal importance of the approximate character of the classical mechanics. In a complex situation approximations may play a decisive role because classical chaos leads to sensitivity to small corrections which impact accumulates. In particular, if these corrections have a  certain "informational direction", then the resultant global picture may differ qualitatively from the one expected on the basis of mechanics. (The limitations of the classical mechanics perspective on chaos
are discussed qualitatively in Sec. \ref{ChapFive} and quantitatively in Secs. \ref{ChapTwelve}-\ref{ChapThirteen}.) It is natural to consider the corrections introduced by the quantum mechanics, but these, at least at the moment, have not led to a significantly new understanding of the questions posed above.

The fact that the information on the physical world provided to us by the classical or quantum mechanics does not say essentially anything about the order in chaos leads one to think that in the conventional approach we miss something fundamentally important, something that may play the critical role in the structure of chaos and thus of our world.  We are lacking general principles of orientation in complex situations and that's why the order that often arises in those situations looks to us accidental. The criticality of the posed question is manifested by the almost complete absence of the general guiding principles allowing to get oriented in situations far from equilibrium. And at the same time, the studies of complex phenomena, and the very existence of different areas of knowledge with different principles of operating with objects, create an impression that we are surrounded by the unknown principles of organization of chaos, the approach to the systematization of which is absent today. Our purpose in this article is to suggest such an approach.

The following paper is organized as follows. Its first part consisting of Sects. \ref{ChapTwo}-\ref{ChapFive} is devoted to the
principles providing background for the theory exposed in the second and third parts.
Secs. \ref{ChapSix}-\ref{ChapTen} describe the general structure of the Heds-Web approach which basic notions are introduced. Then
Secs. \ref{ChapEleven}-\ref{ChapFourteen} discuss the question of emergence of spatio-temporal relations from the purely informational
"Heds-Web". In Conclusion we
summarize the results and discuss some questions that the Heds-Web hypothesis could help resolving.

\section{Possible incompleteness of the traditional approach considering space and time as primary notions}
\label{ChapTwo}

Both classical and quantum mechanics are built on the basis of the postulate that space and time are fundamental, primary elements of reality. Thus building our theories on the basis of space and time we from the very beginning exclude some possibilities in nature that may be precisely the ones determining the order of chaos in reality.  The postulate presumes that all that exists is subordinated to the spatio-temporal order. In particular, all possible interactions of objects should be subordinated to this order and admit an adequate description in spatio-temporal terms. By this, from the very beginning, quite a strong assumption is made on the types of possible interactions in nature. A critical examination of this assumption demands a study of the consequences of the hypothesis that nature has a wider structure. (It should be stressed that relative self-consistency of some approach to reality does not say yet anything about the reality itself, besides that this approach does not contradict it. In particular, it is not excluded that reality may allow several different, relatively self-consistent approaches at the same time.)

A consistent study of the question if space and time really constitute the all-encompassing, universal order, reigning unconditionally, at least, in the usual everyday situations, has received relatively little attention until now.
Yet, as such, the classical Newtonian outlook at space-time as a rigid, unshakable structure of reality has undergone significant changes. The general theory of relativity led to the view of space-time as a dynamical and flexible structure which geometry is not fixed a priori. The theory, however, continues to accept space and time from the very beginning as the exhaustive order fit by everything happening in nature. In turn, some applications of the quantum mechanics to the study of space showed that the latter may be not the basic element of the theory, but, rather, an entity emerging on the basis of interactions of primary discrete objects. However, here as well, the considered fundamental structures should from the very beginning obey the demand to reproduce space-time as the universal order of nature, at least, in the situations experimentally achievable today.

It should be remembered that space and time are just our ways to introduce logic in the natural phenomena. The logic of the nature itself may be more complex. If there are phenomena subordinated to more complex types of order than the spatio-temporal one, then the logic based on the latter, is incomplete. Today, natural phenomena are considered starting from the assumption that the spatio-temporal order exhausts all possibilities of nature. In particular, the order taking place in complex situations is considered as emerging largely accidentally from the interaction of a large number of elements in space and time. Thus, from the very beginning by limiting our viewpoint by space and time we close for ourselves the possibility to consider those properties of nature that are not describable in spatio-temporal terms. Just such properties may be possessed by complex phenomena in chaos. Thus a necessity arises to have a different, wider approach allowing to take into account the possible presence in nature of various types of order.

Thus the search for order in nature suggests the reconsideration of the postulate that space and time are primary notions. In the approach to this reconsideration that is suggested below, one speaks of order and connections not contradicting the spatio-temporal ones, but extending them.

\section{Resonant Web principle and the spatio-temporal section of reality}
\label{ChapThree}

The Resonant Web ("Heds-Web") approach, which is the subject of the second part of the paper, allows to build a physical theory not introducing the demand that all interactions of objects should from the very beginning be subordinated to the spatio-temporal order. Within the frame of this approach space and time are not the primary elements of the theory. The choice of the primary elements appeals to the fundamental, basic principle - the nature consists of interacting objects. Any real situation or experiment include complex objects (remind that waves may also be considered as consisting of objects) and their interactions. These are the latter which are the primary elements of the Heds-Web principle.

The consideration of objects and their interactions as primary elements of the theory is essentially the return to the approach to reality preceding the creation of any system of thinking about it, with the purpose of creating a wider system. From the very beginning, in our approach to reality, the objects exist as such. The objects interact and these interactions happen along certain "tones", in accordance with the inner characteristics of these objects. In addition, an object is from the very beginning limited by the types of interactions in which it is able to participate. It makes sense to consider a certain Order - a system of measurements establishing relations of a certain type between objects and convenient for describing the given "section" of reality.

The order on which physics is based is the spatio-temporal order (below called also "classical"). Such an order is largely anthropocentric, being strongly connected with the task of the local survival of human in space and time. It is good for describing some interactions and it may turn out to be incomplete for describing others. The system of measurements connected with the classical order are measurements of distances between objects with the help of rods, that prescribe objects spatial configuration, and the correspondence between changes of those configurations and readings of physical clocks.

In the light of anthropocentric nature of the classical order, it can be expected that in nature there are interactions that are not "caught" with the help of measurements based on rods and clocks and their different extensions. We suppose that these interactions lead to the randomness that we observe in chaos from the viewpoint of
the "spatio-temporal section" (classical chaos is deterministic and "randomness" here means effective randomness of a single realization
of dynamics). Including these interactions within the structure of physics (and thus necessarily going out of the frame of the spatio-temporal section of reality), we have a chance to consider randomness as a law.

\section{Heds}
\label{ChapFour}

Thus, on the basis of what was said above, experiments based on the spatio-temporal order could miss a kind of interactions that we call "informational resonances" or heds (from the Hebrew "hed" - response, echo). Hed is an interaction of objects that does not belong to the spatio-temporal order, that is an interaction that cannot be measured with the help of rods, clocks or their extensions. Informational resonance is in principle not describable in the spatio-temporal coordinates, constituting a purely informational interaction of objects. Belonging to a different order, hed should not decay with the spatial, and thus according to the Lorentz-symmetry of space-time, also with the temporal separation of objects. The name "informational resonance" or "hed" is connected with this special property of arbitrarily separated objects to be "tuned" to each other. As it will be discussed in the next chapter and more in the second part of the paper, this does not mean that heds do not influence the spatio-temporal relations.

Today the main indication of the possibility of existence of interactions of this kind is the quantum entanglement. For concreteness consider the example of two separated particles with spin one half, in the singlet state. Measuring spin of one of the particles along some direction in space, the second particle instantaneously (at least, on the level of the formalism) starts to have spin opposite to the measured one. Following literal interpretation of the formalism, it can be assumed that a real interaction of two particles happened instantaneously on the level of pure information, in this case the quantum one. Because the distance is unimportant here (quantum entanglement does not decay with the distance) then, considering the viewpoint of the moving observer with the account of the Lorentz symmetry of space-time, we conclude that interactions of this kind may involve objects that are arbitrarily separated from each other both in space and in time. It can be said that the objects are in an informational connection independent of the spatio-temporal separation. Speaking from a more general viewpoint, the quantum mechanics unambiguously indicates the possibility that behind it there is a theory with non-local interactions.

Objects that resonate "information-wise" are in the state of knowledge of each other. We use the term "knowledge" in a concrete meaning: for example, two entangled particles have a certain type of knowledge of each other because whatever happens with one of the particles, the second one immediately "gets to know" about this, being in a hard connection with the second particle.

Below we will assume that the quantum entanglement corresponds to a certain hed of objects and we will use its known properties to describe properties of heds that appear general. In the frame of this assumption a most important, fundamental property of purely-informational interactions becomes clear, namely their indivisibility - a complex system participates in a hed, as a whole. In the example of entanglement considered above this property can be seen considering systems instead of particles. Hed involves complex, spatially extended systems instantaneously and completely.

The assumption that complex systems participate in heds principally as unified, indivisible objects is in agreement with the characteristic features of the phenomenological description of nature. In different natural situations particular "elementary particles" - "indivisible objects" arise, pertinent to the considered situations. These "particles" are by themselves complex and consist of more elementary units organized into a unified whole to form that particle. We will call such an "elementary" particle a "creit" (from "create" and "crate"). Creit is a complex object that can be considered structurally stable in the considered situations, where its "abilities" to participate in heds may be considered constant (properties of creits are considered in more detail in Sections VI-VII). Let us add that wholeness of heds leads to the necessity to consider various situations on the whole, well known in the quantum mechanics. 

\section{Chaos as the order of heds}
\label{ChapFive}

Thus, in the Heds-Web approach, the spatio-temporal relations are considered not as an inherent part of nature, but just as a way of measuring the latter. In contrast, the objects interactions are considered as an inherent part of nature. At the same time,
in many situations it is possible to define the spatio-temporal relations in a consistent way and find out that they obey approximately the connections described by the laws of the classical mechanics. In those special cases where there is no chaos in the system, the approximate character of the laws of mechanics is not so important and  the mechanical connections play a dominating part in the connections of the system objects. In particular, this is the situation of two point objects with a central force interaction. This situation, to a large extent, is at the origin of the notion that in the interaction of two objects the spatio-temporal relations are a closed, self-explaining entity that does not allow for any additional influence besides the laws of mechanics. In reality, however, the objects are usually complex and besides special cases the system is chaotic and the laws of mechanics bring little information.

It is in a chaotic situation that heds play a decisive role. Their influence is realized via small corrections to the laws of classical mechanics
that arise due to the incompleteness of the spatio-temporal section of reality, see Sec. \ref{ChapThirteen}.
These small corrections, being amplified by chaos, lead to a significant global change and,
in the end, due to their stable informational direction, it is them who fix the resultant global structure of chaos.
The latter may differ radically from the one expected on the basis of the locally applicable classical mechanics.
The arising picture of chaos is a complex order where the global structure is determined by quite stable influences of heds of different types.
Such a picture differs strongly for the traditional one where chaos is considered as disorder, while organization in chaos appears to be largely
phenomenological.

It should be stressed that the order of heds is a complex order which is no simpler than the order described by the quantum mechanics. Consider as an example a hed of two creits. Resonating creits are in tune with each other, though the logic of this tuning may be very complex. In particular, from the viewpoint of a classical observer (observer in space-time) the objects behavior may look random. Nevertheless, the creits are in informational connection so that "randomness" has its logic - the logic of the informational resonance. The latter leads to hardly identifiable patterns and tendencies in chaos, subtle correlations and a kind of coordination in the behavior, in short - introduces non-randomness in chaos. This creates the principal possibility of a statistical description akin to the one that emerged in the quantum mechanics.

\section{The world as a complex information web}
\label{ChapSix}

We pass to the consistent construction of the theory. The Heds-Web approach is a purely informational approach to physics that aims to construct
physics starting from purely informational concepts of objects and their informational connections - heds.
We suggest that from the viewpoint of pure information, an object is an entity possessing information like 
in the ordinary view an object is an entity possessing energy.
Thus an object ("creit") is essentially, a certain "crate" of pure information.


Creits create heds-webs by entering into heds. Within the Heds-Web approach, the world constitutes a giant resonant Web - a set
of objects unified by purely informational connections into a single general structure. This Web is in the state of a constant
flux - an infinite, incessant change of connections, happening everywhere and in all directions (in the sense of flux, this state
is most close to the state of quantum fields, however it is not random and it does not happen in space and time. Thus a new category
is needed). Any attempt to "stop" this flux and "seize" the state of the Web, thus decomposing the flux into the "current state" and
its changes introduces an inaccuracy, both qualitative and quantitative. Besides, a linear sequence of causes and effects, prescribing
a certain direction of the flux, is inapplicable for the Heds-Web (remind that heds are non-local in time). In this paper we
will consider an approximation within which one can conditionally single out "pieces" of the Web and speak of the structure of these
pieces and the processes of their change. Let us note that speaking of these processes it makes no sense to speak of their temporal
duration - time is a way of measurement connected with matter and it does not apply to the processes occurring at the level of pure
information. 

If it is necessary to perform a fundamental consideration of the structure of reality, one may choose as elementary constituents of
the Web sufficiently stable
elementary particles, such as those appearing in the standard model or more fundamental. The determination of stable constituents
presents no difficulty in the practically important situations, that are of main interest to us here.

In the Heds-Web approach the spatio-temporal relations between objects are considered as metric characteristics of the Web that
are resultant from heds. In particular, the interactions of objects in space and time that are nothing more than the approximate
description of the logic of changes of spatio-temporal relations between objects, are also considered as resultant from heds.

\section{The general structure of the theory}
\label{ChapSeven}

The formalism described
below is based on the maximally economic, in our view, assumptions on the structure of the web and the flux that include the following
propositions:

-connections can be of different types - tones

-connections can be more strong or less strong

-entering into hed is a threshold phenomenon (this is necessary to define a non-trivial dynamics, otherwise all creits
will immediately enter in heds and by this the dynamics will end. The threshold nature of heds is also important for the
emergence of a natural mathematical structure to describe transformations of pure information in heds, see below)

-a web formed by resonating creits is able to act in the world of pure information as a whole. This is demanded by consistency
because creit is a complex object and thus is also a web, see Fig $1$. The web properties as a whole are a non-trivial function of the informational characteristics of the creits comprising the web.

\begin{figure}[ht]
\vspace{0.5cm}
\begin{tabular}{cc}
 \epsfxsize=8.5cm  \epsffile{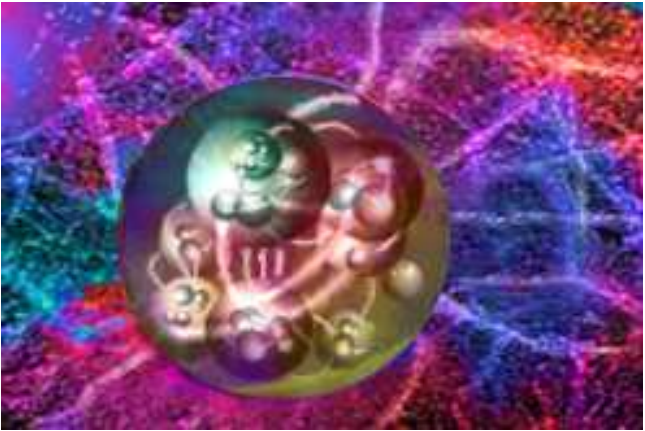}\\
\end{tabular}
\vspace{0.5cm}
\caption{Shown is a depiction of a creit. The picture shows that the creit can also be considered as a web,
illustrating the principle that whether an object should be considered as a creit or as a web depends on
the scale of the consideration.} \label{eps0.8}
\end{figure}

On the basis of these basic, and some additional, propositions, a very non-trivial dynamics arises with many combinatorial possibilities. This dynamics is, in our view, able to describe all structure of nature, including chaos and quantum mechanics.

\subsection{Hed tone}

It is natural to assume the possibility of the existence of different types of heds that we will call tones. A tone is a certain
type of pure information that is involved in the hed. In the frame of this article the exact number of existing tones (from one to
infinity) is not essential.

\subsection{Creit characteristics}

As already mentioned, creit is a "crate" of pure information. Correspondingly each creit can be characterized by
a unique set of numbers $I_k>0$ that give an amount of information that corresponds to the tone k (we will assume $I_k\neq 0$ for all
k). The amount of information $I_k$ determines the intensity with which the creit enters into heds according to tone k, and thus, in
principle, $I_k$ can be measured by considering the hed of the considered creit with a creit chosen as standard. Eventually it is the
set of numbers $I_k$ that determines the creit behavior in a complex, chaotic situations and decides the general nature of the creit
interaction with the environment.

\section{The simplest informational process: hed of two creits}
\label{ChapEight}

Having introduced in our view the most natural characteristics of creits, let us consider their simplest informational interaction
which is a hed of two creits.

\subsection{The connection strength and the hed threshold}

As mentioned above we will assume that resonant connections may be more strong or less strong. The measure of the connection
strength is the overlap - an index of proximity of two creits (often instead of a distance as a measure of difference of two
objects, it is more natural to introduce the so-called overlap, see e. g. \cite{RTV}). The overlap $F(x, y)$ is a symmetric
positive function of two, in this case, positive quantities. It allows to calculate the strength of the creits hed according to the
tone k so that if one creit has $I_k=x$ and the other $I_k=y$ then $F(x, y)$ is the strength of their hed according to tone k. Here we assume
that the function $F(x, y)$ is the same for all tones, characterizing some universal type of "distance" for information spaces.

Furthermore, as mentioned above, entering in hed should be a threshold phenomenon. We assume that the function $F(x, y)$ is chosen
so that the threshold condition according to tone k has the form $F(x, y)=1$. Then for $F(x, y)<1$ the hed according to tone k
is not possible for the considered two creits, while for $F(x, y)\geq 1$ the hed occurs. Then, if for all k we have $F(x, y)<1$ then no hed arises between the two creits (thus,
not any two creits can enter into hed). Otherwise a hed starts according to all tones for which $F(x, y)\geq 1$ (it can be one or
several tones, see Fig. $2$). Two creits may enter into hed only if both have a sufficiently pronounced common tone.

\begin{figure}[ht]
\vspace{0.5cm}
\begin{tabular}{cc}
 \epsfxsize=8.0cm  \epsffile{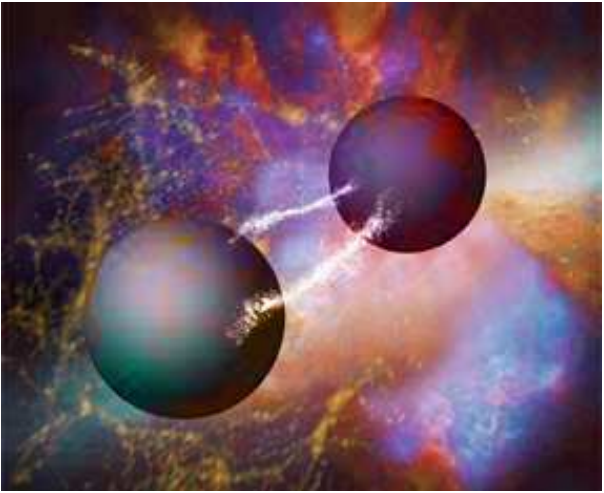}\\
\end{tabular}
\vspace{0.5cm}
\caption{Two resonating creits with the Heds-Web in the background. The hed occurs according to two different tones. The shorter connection
is the strongest and is the dominant tone of the considered hed.} \label{eps0.8}
\end{figure}

It should be noted that similarly to the quantum entanglement and disentanglement, hed happens or does not happen instantaneously, involving
the two creits in a unified and indivisible way. The impossibility to sort out which of the creits is the reason for entering into the hed
(the absence of hierarchy) reminds of an instantaneously occurring non-linear process where it is not possible to distinguish the action and
the response. Otherwise said, in the hed it is not possible to single out the "initiating" and the "responding" creits, and it is only possible to speak of an interdependent passage to the state of being connected. These properties are necessary so that heds could really describe all the complexity of chaos.
The instantaneous nature of the hed from the viewpoint of the physical time is connected with the fact that the hed, as a phenomenon,
does not happen in space-time.

Tone for which the hed strength is maximal will be called the dominant tone of the hed. The dominant tone determines the hed essentially when 
the hed happens according to several tones at once, by playing the part of the dominating informational
principle connecting the creits, see below the discussion of the law of the dominant.

\subsection{Two resonating creits as the simplest web}

Two resonating creits form the simplest heds-web. Because creits by themselves are complex objects then consistency demands that a web is
able in its turn to enter the informational interactions as a certain new effective creit (the difference of a web from a creit is only in
that in the considered processes the web of a creit is assumed unchanged). This means that the web acquires its own values of amounts
of information according to which the hed happens. With respect to the tones for which there is no hed between the creits, the latter act independently. A new function
$I(I_1, I_2)$ arises characterizing the hed, that gives the amount of information of the web of two resonating creits with original
amounts  of information $I_1$ and $I_2$ (and thus satisfying $F(I_1, I_2)\geq 1$). This function should not be a simple sum of the amounts
of information of the resonating creits, because this would mean that the system of two resonating creits behaves simply like the two creits
separately (this is analogous to the fact that entropy of a composite system consisting of two independent systems is the sum of the entropies
of the systems. Let us note that the connection of pure information and entropy demands a further study which is beyond our purposes in this
work. Here it is only appropriate to say that this connection does not look simple). It appears evident that the sign of the difference
$I(I_1, I_2)-(I_1+I_2)$, expressing the qualitative meaning of the notion of hed, should be the same for all $I_1$ and $I_2$ obeying the
threshold condition $F(I_1, I_2)\geq 1$. We postulate the super-additivity that is the inequality
\begin{eqnarray}&&
I(I_1, I_2)-(I_1+I_2)\geq 0,  \label{basic5}
\end{eqnarray}
that, roughly speaking, expresses that the web, besides the information on the original creits, carries the information that they entered in the
connection.

Super-additivity reflects the qualitative difference between resonant and mechanical connections. Mechanical connections constitute constraints on
the possible motions of the objects. Heds are the opposite of constraints, they endow each creit with greater "flexibility and mobility", expanding
their possibilities beyond the mere sum of the original possibilities of the resonating creits. This is the meaning of the inequality above. It can  be illustrated using the example of the quantum entanglement. The entangled objects sharing the joint wave function may participate in additional effects
of quantum interference in comparison with the situation where their joint wave function is a simple product of the wave functions of each object
separately, that corresponds to the disentangled state.

Super-additivity expresses the above picture quantitatively because it means the information in the hed is not merely summed but it is "multiplied". Indeed, for positive $x$ and $y$ the usual multiplication is super-additive provided the threshold condition
\begin{eqnarray}&&
\frac{xy}{x+y}\geq 1,
\end{eqnarray}
is satisfied. Thus $I(x, y)$ and $F(x, y)$ are parallel to $xy$ and $xy/(x+y)$ correspondingly.

\subsection{Field-Space}

The fact that entangled objects share a joint information field of the wave function to which the entangled objects "belong" may be
considered as a counterpart of an important notion of the Heds-Web: the Field-Space.

The increase of the informational space of "possibilities" thanks to the hed, as described by the inequality (\ref{basic5}) admits a simple
interpretation that resonating creits share a joint "space" created by their hed where the space "volume" $I(I_1, I_2)$ is such as to contain
the two creits with their original volumes $I_1$ and $I_2$. The structure of the new effective "creit" constituted by the web describes a
qualitative change in information which amount increases formally thanks to the hed. This emerging structure plays a principal role in the Web transformations
and it will be called the "Field-Space". The Field-Space constitutes a unified information field of the resonating creits. The latter are inside the
Field-Space so it has the characteristics of space. On the other hand, because  the web has features of an  effective creit, then a creit
which is outside the web will first enter in informational interaction with the web as a whole and not with the creits forming the web.
Thus this creit will interact with the Field-Space similarly to how particles interact with a field, so that the Field-Space has characteristics
of a field.

Field-Space, naturally, arises for any web, not just the one formed by two creits. The building up of the Field-Space by creits and
their connections reminds of the connection between the "skeleton" and the space, used in the Regge calculus \cite{MTW}, and also the building
of space or space-time
in some theories
of quantum gravity on the basis of graphs, though the difference is  significant. In contrast to the more usual space,
the Field-Space has a sharply defined boundary - it includes the creits of the web and does not include the creits outside the web. Let us note
that, considering the world as a giant connected Web, we come to the conclusion of existence of a purely informational unified Field-Space of the
world.

\section{Properties of the Heds-Web Structure}
\label{ChapNine}

In the previous chapter we considered the main characteristics of the resonant connection
on the example of a hed of two creits. As the number of creits participating in the heds increases, the complexity
of the emerging webs grows fast. For illustration consider the case of three creits, two of which form a web.
Then the threshold for the third creit to enter into hed with the Field-Space of the web is lower than the threshold
with each one of the web creits separately. There are four ways for the third creit to connect to the web: to one creit, 
to another, to each one of them and to their connection. The latter way is a new one in comparison with the case of 
two creits, it corresponds to the formation of hed with the Field-Space of the web of the two creits. 
In all these cases a new web arises with its own unique Field-Space. At further increase of the number
of creits forming the considered webs, the number of possible combinations grows fast. In connection with the
large number of combinations and the non-trivial laws of transformation of the amount of pure information at the
formation of heds, the emerging multi-tone webs possess large complexity
(see Fig. $3$ for an illustration) leading to a great capacity of information "encoding" in such webs.
\begin{figure}[ht]
\vspace{0.5cm}
\begin{tabular}{cc}
 \epsfxsize=8.5cm  \epsffile{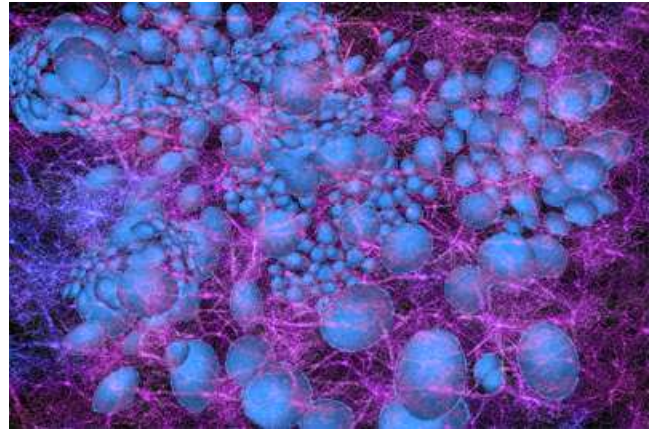}\\
\end{tabular}
\vspace{0.5cm}
\caption{Illustration of the Heds-Web of many creits. The creits amounts of information provide a natural volume measure.} \label{eps0.8}
\end{figure}

While the detailed consideration of the emerging web structures is beyond the frame of this article, their general character can be
realized empirically as follows. It is a universal property of nature to produce separation of "scales of information", leading to
the existence of approximately closed systems of objects that interact among themselves much more intensely than with the environment.
In the frame of our approach such picture means that every "macro-piece" of the Web consists of many intertwined, interdependent webs,
where a conditional sub-web may be singled out by separating the scales of information (we consider the strengths of the connection between
creits, as measures of their proximity, cf. Fig. $2$). One can imagine a structure similar to the structure of the
visible universe where stars group into galaxies, while galaxies into clusters of galaxies. Then from the viewpoint of the scale of the
galaxy, the latter is a web, while from the viewpoint of the scale of the cluster - the galaxy behaves approximately as a unified whole
or an effective creit.

We expect creits to have their own "informational strategy" or directionality of the informational connections into which they enter.
This assumption is connected with the expectation that the extreme complexity of the Heds-Web described above should lead to great sensitivity of the
the Web to the properties of the creit, that is one can expect to have a rather robust correspondence between the creit properties and
its connections. The quantitative expression of this expectation is "the law of the dominant".

Out of the many connections in which a creit finds itself one can distinguish the dominant one, - the connection which strength is
maximal. The dominant connection significantly determines the rest of the creit connections because other connections exist on the
background of a more information-saturated Field-Space created by the dominant connection. This can be seen by a thought experiment
considering how the existing web of the creit builds up starting from the dominant connection and continuing with the rest of the
connections. "The law of the dominant" says that the dominant connection introduces an essential orientation in all the rest of the
connections of the considered creit. The importance of the dominant is that it is the dominant that determines what could be called
"the informational strategy". The rest of the connections are relatively subordinated to this strategy, that is to the dominant
connection.

\section{The Properties of the Heds-Web dynamics}
\label{ChapTen}

The dynamics of the Heds-Web is centered at the assumption that the Heds-Web is in the state of an infinite flux of a constant
change of connections happening everywhere and at every "point" in all "directions". This incessant change of connections is the
essence of the phenomenon we call chaos (let us stress that in contrast to the traditional view, here chaos is the dynamics of the
connections and not of objects).

The basic proposition regarding the processes of changes of connections is that all heds that are allowed by the present thresholds
may happen. In particular, creits may enter an indefinite number of connections. The rearrangement of the connections leads to a redistribution
of the pure information in the Web. This redistribution occurs constantly.

One can expect the existence of heds of three different dynamical types (in contrast to static types - tones) that describe different dynamical
possibilities. At a "constant" hed of two creits they are all the time in the hed. A single-time hed corresponds to the situation where
creits enter into hed, that after producing its effects ends and the creits do not enter in heds any more. Finally, a repeating (pulsating)
hed corresponds to the situation of repeating acts of entering into the hed and ending it.

In the Heds-Web approach chaos is not a disorder, but to the contrary it is an extremely complex and definite order of changes of connections
in the Heds-Web ("the world is chaotic but not arbitrary or accidental"). This order is significantly determined by "the law of changes of
dominants" that is the dominant connections. Because it is the dominant connection that orients all connections of a creit, then significant
rearrangements of the Web occur just at the changes of the dominants (entering into a connection stronger than all the existing ones). Each
such change is a "butterfly effect" that is a change of one connection in the Web that leads to changes of many other connections. Let us
note that in the Heds-Web approach due to the presence of definite elements - connections, the notion of the "butterfly effect" has a definite
meaning.

\section{Spatio-temporal relations as resultant from heds}
\label{ChapEleven}

In the frame of our hypothesis the spatio-temporal relations constitute a certain description or an observable section of the Heds-Web. 
These relations are a certain manifestation of full informational connections between the creits of the Heds-Web. We first consider 
the spatial description of the Heds-Web. 

The existing Heds-Web, consisting of creits and connections between them, leads to the existence of a certain spatial configuration of 
objects in the classical order known to us. The spatial distance between two given objects is determined not only by the connections 
of the corresponding creits but also by the whole picture of connections in the Heds-Web. It appears sensible to assume the existence of a 
positive correlation between the strength of the creits connection and their spatial proximity, manifested depending on the 
whole set of connections in the Heds-Web. 

A convenient way for describing the correspondence between the Heds-Web and the spatial configuration of objects is the consideration 
of the correspondence between the distribution of information in the Field-Space and the distribution of the mass-energy in space. The 
latter constitutes quite a complete description of the spatial configuration where objects are represented by a sharp increase of the 
mass-energy density. Analogously, the distribution of information characterizes creits and their closeness in the sense of the strength 
of their connection. 

In accord with the above we introduce the first postulate of the correspondence. 

1. Correspondence between Field-Space and space (the latter understood as a space-like hypersurface in space-time):

The distribution of information in the Field-Space of the Heds-Web is manifested in space as a certain distribution of energy. Thus, 
the mass-energy density $\rho$ appearing in the Einstein equations reflects the distribution of information. 

Let us remind that our general approach is aimed at considering what stands behind space and time so to achieve a more complete 
view of the latter and the laws described within their frame. Correspondingly we will assume that the space-time emerging from the 
Heds-Web is described by the general theory of relativity, that is by a metric satisfying Einstein's equations. This leads to the 
second postulate connecting the distribution of energy,  reflecting information, and the space-time geometry. 

2. Correspondence between the energy distribution on space-like hypersurfaces and the space-time metric:

This correspondence is based on the demand that the Cartan moment of rotation and the density of mass-energy are equal for each space-like hypersurface according to

\begin{eqnarray}&&
R+ (Tr {\bf K})^2- Tr K^2=16 \pi \rho, \label{basic1}
\end{eqnarray}
where $R$ is the scalar curvature invariant of the $3-$geometry intrinsic to the hypersurface, while
$K_{\alpha\beta}$ is the extrinsic curvature tensor.

Here and below we follow the notations of \cite{MTW}. As it is well known the second postulate leads
to the Einstein equations
\begin{eqnarray}&&
G_{\alpha\beta}=8\pi T_{\alpha\beta},\label{basic2}
\end{eqnarray}
where $G_{\alpha\beta}$ is the Einstein tensor and $T_{\alpha\beta}$ is the stress-energy tensor. The correspondence
is realized as follows. Designating the local $4-$vector normal to an arbitrary spacelike slice through spacetime
by ${\bf u}$, one has
\begin{eqnarray}&&
\rho=u^{\alpha}T_{\alpha\beta}u^{\beta},\\&&
R+ (Tr {\bf K})^2- Tr K^2=2 u^{\alpha}G_{\alpha\beta}u^{\beta}.
\end{eqnarray}
Then Eq.~(\ref{basic1}) leads to
\begin{eqnarray}&&
u^{\alpha}G_{\alpha\beta}u^{\beta}=8\pi u^{\alpha}T_{\alpha\beta}u^{\beta},
\end{eqnarray}
which is equivalent to Eq.~(\ref{basic2}) by the arbitrariness of the spacelike slice.

The main new component of the suggested approach to the space-time is the outlook at the energy distribution as described by the 
postulate 1. The questions of what the energy is and how it is connected with the notion of information are usually
outside the frame of physics. In the suggested approach the mass-energy distribution in space reflects the distribution of information 
in the Field-Space created by the creits and their connections. (It should be added that the mass-energy
density above becomes meaningful only after the metric is found.)

Even though the second postulate is one of known ways of establishing a connection between the energy distribution and the
metric, it acquires a new qualitative meaning from the viewpoint of heds. Because the mass-energy density reflects the 
distribution of information created by creits and their heds, then Eq.~(\ref{basic1}) says that creation of additional heds between 
given creits leads to additional curving of the space-time. For example, when two creits enter into hed some additional information 
appears (because of the super-additivity) that leads to the change of the energy distribution and eventually the Cartan moment 
of rotation. Thus, heds influence what happens in space-time by changing the moment of rotation (the corresponding influence on 
chaos is discussed in the following two Sections).

Postulate 1 also gets an additional meaning. 
Because every creit in the Heds-Web, generally speaking, participates in many connections, then it can be said
that the Heds-Web possesses a certain generalized "elasticity". The existing connections "resist" the creation of
new connections in accord with the law of the change of the dominant. Then the parallel
between the theory of curved space-time and the elasticity theory, pointed out by Sakharov \cite{Sakharov}, acquires
a new meaning - the Einstein equations express the metric aspect of the Heds-Web elasticity. Let us note however that
the parallel with the Sakharov view, who spoke of elasticity of space, is but qualitative - in our approach
the space does not exist by itself and the Heds-Web is not embedded in space in contrast to the quantum vacuum fluctuations.
Also the Heds-Web does not constitute a "skeleton" for space of a kind discussed in the Regge calculus \cite{MTW} (rather
it a "skeleton" for the Field-Space).

\section{The view of the laws of classical physics as a consequence of empirical constitutive relations for the stress-energy
tensor}
\label{ChapTwelve}

The Bianchi identity $G^{\mu\nu}_{;\nu}=0$ allows to obtain from the Einstein equations (\ref{basic2}) the law of energy-momentum conservation
\cite{MTW}
\begin{eqnarray}&&
T^{\mu\nu}_{;\nu}=0. \label{basic4}
\end{eqnarray}
Thus, in the Heds-Web approach it appears natural to consider the law of energy-momentum conservation as a constraint resultant from the use of
the spatio-temporal coordinates to describe pure information. It is known that \cite{MTW,Infeld} the laws of motion of the
classical physics can be obtained from Eq.~(\ref{basic4}) by prescribing a correct constitutive relation for the tensor
$T^{\mu\nu}$. For example, the Maxwell equations can be obtained by substituting into Eq.~(\ref{basic4}) the constitutive
relation
\begin{eqnarray}&&
T^{\mu\nu}=\frac{1}{4\pi}\left(F^{\mu\alpha}g_{\alpha\beta}F^{\nu\beta}-\frac{1}{4}g^{\mu\nu}F_{\sigma\tau}F^{\sigma\tau}\right),
\end{eqnarray}
where $g_{\alpha\beta}$ is the metric and $F_{\mu\nu}$ is determined by the $4-$potential $A_{\nu}$ via $F_{\mu\nu}=\partial_{\mu}A_{\nu}-
\partial_{\nu}A_{\mu}$. Classical particle dynamics may also be obtained in this way \cite{Infeld,MTW})

We see that the above derivation of the Maxwell equations does not differ in essence from the derivation of
the equations of hydrodynamics where the energy-momentum conservation is considered as given and then a constitutive relation for
$T^{\mu\nu}$ is prescribed (the particle number conservation is also added as a rule). In particular, the field $A_{\nu}$ plays
the role of a "slow" field or the order parameter. As it is well known in hydrodynamics, the used expressions for $T^{\mu\nu}$ are 
effective approximate relations.
In accordance with this, it seems natural to consider the possibility 
that the laws of physics are effective,
empirically confirmed descriptions of the stress-energy tensor, that provide for a well-working locally in space and time approximation
to the dynamics of energy. That is, guided by the spatio-temporal logic, good, self-consistent local approximations for
$T^{\mu\nu}$ are achieved in terms of some effective variables which determination is the subject of the research. It is these approximations that
constitute the laws of physics.

Such a non-absolutist view may be very useful for approaching
such problems of the usual formal approach as for example the
"infinite self-energy" of a point particle (the decomposition of $T^{\mu\nu}$ into the field and the particle becomes the question of
finding an effectively working description). It is more important to us here that this demonstrates the existence of corrections to the known expressions for $T^{\mu\nu}$ and thus to the equations of motion. Let us introduce the decomposition
\begin{eqnarray}&&
T^{\mu\nu}=T^{\mu\nu}_K+T^{\mu\nu}_U,
\end{eqnarray}
where $T^{\mu\nu}_K$ is the component of the tensor the expression of which is known from the laws of physics, while $T^{\mu\nu}_U$ is
the correction existing due to the existence of the more fundamental description. Formally this component can be defined via (let us note the
similarity of the used procedure with the one used in the analysis of the dark energy, see e. g. \cite{Starobinsky})
\begin{eqnarray}&&
8\pi T^{\mu\nu}_U=G^{\mu\nu}-8\pi T^{\mu\nu}_K.
\end{eqnarray}
Then the usual laws of motion that would follow from $(T^{\mu\nu}_K)_{;\nu}=0$ acquire two corrections - one due to the correction to the
metric produced by the term $T^{\mu\nu}_U$ in the Einstein equations, while the other is due to the appearance of an additional term in the
RHS of the equation $(T^{\mu\nu}_K)_{;\nu}=-(T^{\mu\nu}_U)_{;\nu}$ [here it makes no sense to analyze where one can shorten the equation to
$(T^{\mu\nu}_K)_{;\nu}=(T^{\mu\nu}_U)_{;\nu}=0$]. In the case where the solutions of the original equations following from
$(T^{\mu\nu}_K)_{;\nu}=0$ are sensitive to the corrections, the resulting picture obtained with the account of the latter may differ
qualitatively from the one expected on the basis of the "spatio-temporal" approximation $T^{\mu\nu}\approx  T^{\mu\nu}_K$. According to our
assumption this is just the case of the chaos.

\section{Heds-Web and Classical Mechanics}
\label{ChapThirteen}

Thus, according to the Heds-Web hypothesis, the equations of the classical mechanics must
have small corrections due to $T^{\mu\nu}_U$ (not only of quantum nature). From the viewpoint of the spatio-temporal section, these corrections are illogical.
The apparent absence of logic in them is determined by the discrepancy of the spatio-temporal description and the true logic of chaos - the
logic of heds. From the viewpoint of a classical observer, it would be natural to model these corrections as random which
is just what is effectively done in chaos where a statistical description is introduced almost necessarily. In particular, such explanation
of the effective randomness could lead to further understanding of the foundations of the statistical mechanics and of the reason why the introduction of noise in the equations is so effective. 

A classical observer sees the spatio-temporal results of the heds without seeing their cause. These
results obey approximately the laws of mechanics which are then considered to be their causes. According to the Heds-Web hypothesis 
the mechanics in its approximations misses the essence of chaos as a complex order of the Heds-Web. The amplification 
of the locally small correction due to $T^{\mu\nu}_U=T^{\mu\nu}-T^{\mu\nu}_K$, that distinguishes the Heds-Web order from the mechanical disorder, leads to a qualitative difference of the global picture from the local one.

In fact,  as it was shown in the previous Section, the laws of mechanics (as the laws relying on the energy-momentum conservation) constitute rather a geometrical limitation due to the choice of the way to consider chaos, than the law of chaos. It is this fact that in our view stands behind the 
fact that mechanics brings little information when chaos is considered. In other words, the laws of mechanics look like constraints on the manifestation of heds related to the introduction of the spatio-temporal system of coordinates.
This is also the reason for the big difference of notions of a mechanical and a resonant
connections - one means a constraint, the other means increase of possibilities.

Heds effect on the space-time is indirect - the Cartan moment of rotation characterizing the spacelike slices is their
basic manifestation. The importance of the Cartan moment of rotation for the laws of physics was stressed in \cite{WK}.
From the view-point of space and time heds are manifested via contributions to the energy-momentum tensor that do not
allow for an effective spatio-temporal description (forms of energy not allowing an effective embedding in the space-time).
This clearly reminds of the dark energy the consideration of which,
however, is beyond the frame of this work.

\section{View of the Quantum Mechanics}
\label{ChapFourteen}

The emerging outlook at randomness in the quantum mechanics is quite similar to the outlook at randomness in the classical chaos with a
sole exception that quantum mechanical randomness is considered conventionally as fundamental. From our viewpoint this
fundamentality corresponds to the principal inapplicability of the spatio-temporal order to the consideration of quantum
connections (these connections may manifest heds independently of the spatio-temporal separations as in the example of the quantum
entanglement). The statistics is a description of the 
interaction of a classical observer, acting within the spatio-temporal logic, with a quantum object participating in connections which character 
is not spatio-temporal in principle, taking account for the difference of scales of interacting objects. Let us stress that in the Heds-Web approach there are
no classical and quantum objects, there are only creits participating in that or another connection. The Heds-Web
is a possible objective reality behind the quantum mechanics. The study of the possibility to follow in detail the
emergence of a quantum mechanical description is a subject for future work.

\section{Conclusion}
\label{ChapFifteen}

We considered a possible continuation of the reconsideration of the notions of space and time started by the general theory 
of relativity. The demand not to consider the space-time geometry among the primary elements of the theory was extended to the demand 
not to consider among them the space-time as such. This approach is a natural one for the search of a more fundamental theory 
that could unite the general theory of relativity and the quantum mechanics. We suggested that the theory standing behind the 
space and time is purely informational. As a structure of such a theory we suggested a web - "the Heds-Web". The latter is 
formed by creits (analogs of objects in the "information space" - Field-Space) and their connections - "informational resonances" 
or "heds". The most important consequence of our consideration is a new outlook at the classical mechanics that reveals 
the purely informational character of the order of chaos. This order cannot in principle be caught within the frame of the 
existing approach based on space and time. 

In the paper the foundations of the Heds-Web approach were established and the discussion of laws of transformations of pure 
information when creits enter into heds, was started. The
notion of the Field-Space, as a unified information field of the Heds-Web, was introduced, and the problem of introducing a new
category, connected with the infinite flux state of the Heds-Web, was posed.

Within the frame of the Heds-Web approach, chaos is a definite order of changes of connections in the Heds-Web, that is chaos
in no way means an accident or an arbitrariness (this claim encompasses the quantum mechanics too). The main suggested law
of the order of chaos is the law of changes of the dominants - the strongest connections of the creits. The changes of the dominants
occurring due to the appearance of another, stronger dominant connection rearrange many connections at once, constituting
the "butterfly effect" of the Heds-Web.

We gave a non-contradictory view at the emergence of the laws of classical physics from the pure information. In this
view the energy distribution in space reflects the distribution of information in the Field-Space. The laws of physics 
emerge as effective descriptions of the stress-energy tensor, similar in nature to constitutive relations for the 
stress-energy tensor used in the hydrodynamic approach. While constituting a good approximation locally in space and time, these 
laws may give an incorrect global picture where chaos is present. This is connected with the fact that the approximation used 
in mechanics is too crude to capture completely the informational connections in chaos determining the global picture. In other words, 
there is a qualitative difference between the local mechanical approximation to the laws of chaos and the informational nature 
of the latter. As a result,
mechanics often gives just a self-consistent view that, not explaining anything, describes the result occurring in nature de facto.

The Heds-Web describes the inner structure of chaos as a constant change of connections of objects, but not of objects themselves, 
that remain unchanged. Formally both the Heds-Web and the classical
mechanics say that chaos is determined. However, the classical mechanics is not closed because of the presence of the more fundamental 
quantum-mechanical description. The statistical nature of the latter eventually brings the conclusion that within the traditional approach, based 
on space and time, chaos is fundamentally random.
The Heds-Web approach, by disclosing the structure standing behind the classical and the quantum mechanics, restores the non-randomness 
of chaos. The determinism, however, reemerges at a new level - this is already not determinism in time, but determinism in connections.

Suggesting a structure of the objective reality standing behind both the quantum mechanics and the space and time, 
the Heds-Web is a candidate for the theory of quantum gravity. From the viewpoint of the quantum mechanics, heds play the role of the "hidden" non-local interactions. Being informational connections akin to the quantum entanglement, heds correspond to informational dynamics not simpler than
the quantum-mechanical one. At the same time, both, micro- and macro- objects may participate in heds. That is, within the frame of the suggested picture
of the objective reality, macro-objects may enter into connections not less
complicated than the quantum-mechanical ones. It is these connections, applicable at all scales, that constitute the principles of the 
organization of chaos. Thus, if the Heds-Web hypothesis is correct, then one may expect phenomena not less surprising than the quantum-mechanical ones in the physics of "ordinary" macroscopic objects. In other words, the Heds-Web hypothesis points out to the possibility of principally new ways
of operating objects. In particular, it is not excluded that the principles described in this paper can be used to achieve high-temperature 
superconductivity. 

Let us mention some questions of modern physics which the Heds-Web approach may shed light on. 

The necessity of the statistical description in chaos (where usually
statistics is considered non-fundamental) and in the quantum mechanics, can now be viewed as a consequence of the incompleteness of the
spatio-temporal description of nature.

The properties of complex systems, as a whole, are probably the main characteristics of the organization of chaos that is absent in 
the traditional fundamental approach, where they appear as emerging essentially accidentally. Usually they do not contradict 
the known laws of interactions of parts, but also cannot be obtained from them formally. In the Heds-Web approach, the complex systems, 
constituting resonant webs, possess a larger amount of information than just the sum of the amounts of information of the creits comprising 
the web. Thus, in this approach, the emergent properties of complex system as wholes are fundamental. In view of the fact that 
the emergence of properties of complex systems as wholes is an overwhelmingly widespread phenomenon, the outlook that the emergence is 
a fundamental property of nature appears to us more satisfactory.

Heds, that according to our assumption are as fundamental interactions of objects as the known ones could also shed light at the fundamental questions of physics. In particular, the instantaneous nature of heds could help to resolve some principal difficulties of the modern fundamental theories of matter. In the latter, the assumption of a finite maximal speed of propagation of information leads to the impossibility of considering spatially extended objects as the elementary ones. As a result, one has to consider either point particles or fields determined over continuum that brings about different kinds of divergences.

Summarizing, the reconsideration of the notions of space and time brings about two principally new possibilities that could
be seen while remaining within the frame of the spatio-temporal approach. First, there appears a possibility that chaos has 
an informational nature. Second, there appears a possibility of principally new ways of operating objects. In connection with this, we suppose 
that the possibility of the passage to the purely informational physics, suggested in this work, is of high interest. 



\end{document}